\documentclass[twocolumn,aps,prl,floatfix,superscriptaddress]{revtex4-1}
\usepackage{amsmath,amssymb}
\usepackage{graphicx,float}
\usepackage{times,txfonts}
\usepackage{color}
\usepackage{multirow}
\usepackage{bbold}
\usepackage{color}
\usepackage{soul}
\usepackage{hyperref}
\usepackage{times,txfonts}
\usepackage{natbib}
\usepackage{nicefrac}
\usepackage{blindtext}
\usepackage[export]{adjustbox}
\usepackage{mathrsfs}
\usepackage{textcomp}
\usepackage{blkarray}
\usepackage{multirow}	

\newcommand{\abs}[1]{\left| #1 \right|}

\renewcommand{\epsilon}{\varepsilon}

\def\VR{\kern-\arraycolsep\strut\vrule &\kern-\arraycolsep}
\def\vr{\kern-\arraycolsep & \kern-\arraycolsep}
\usepackage{sistyle}
\usepackage{soul}
\definecolor{lightblue}{RGB}{185,210,248}
\sethlcolor{lightblue}

\begin{document}

\title{Flat Magic Window}

\author{Felix Hufnagel}
\email{fhufn079@uottawa.ca}
\affiliation{Department of Physics, University of Ottawa, 25 Templeton street, Ottawa, Ontario, K1N 6N5 Canada}

\author{Alessio D’Errico}
\email{aderrico@uottawa.ca}
\affiliation{Department of Physics, University of Ottawa, 25 Templeton street, Ottawa, Ontario, K1N 6N5 Canada}

\author{Hugo Larocque}
\affiliation{Research Laboratory of Electronics, Department of Electrical Engineering and Computer Science, Massachusetts Institute of Technology, Cambridge, 02139 Massachusetts, USA}

\author{Fatimah Alsaiari}
\affiliation{Department of Physics, University of Ottawa, 25 Templeton street, Ottawa, Ontario, K1N 6N5 Canada}

\author{Jeremy Upham}
\affiliation{Department of Physics, University of Ottawa, 25 Templeton street, Ottawa, Ontario, K1N 6N5 Canada}

\author{Ebrahim Karimi}
\affiliation{Department of Physics, University of Ottawa, 25 Templeton street, Ottawa, Ontario, K1N 6N5 Canada}

\begin{abstract}
{Magic windows (or mirrors) consist of optical devices with a surface deformation or thickness distribution devised in such a way to form a desired image. The associated image intensity distribution has been shown to be related to the Laplacian of the height of the surface relief. We experimentally realize such devices with flat optics employing optical spin-to-orbital angular momentum coupling, which represent a new paradigm for light manipulation. The desired pattern and experimental specifications for designing the flat optics was implemented with a re-configurable spatial light modulator which acted as the magic mirror. The flat plate, optical spin-to-orbital angular momentum coupler, is then fabricated by spatially structuring nematic liquid crystals. The plate is used to demonstrate the concept of a polarization-switchable magic window, where, depending on the input circular polarization handedness, one can display either the desired image or the image resulting from the negative of the window's phase.}

\end{abstract}	

\maketitle

The mechanism behind ancient magic mirrors from China and Japan was not understood until the 20th century, despite the earliest creations of these artistic pieces dating back to 2000 BC~\cite{berry2005oriental}. The cast bronze mirrors presented as normal mirrors while viewing ones reflection. However, when sunlight was shone directly on the mirror, it acted as a subtly parabolic mirror forming an image---corresponding to patterning on the back side of the mirror---presented on the floor or a screen~\cite{ayrton1879ii}. A similar phenomenon can be observed in the reflection of the sun off large windows and onto a street below. Though the window appears flat and does not significantly distort an image while we look through it, the slight deformations from tension around the edges result in a non-uniform reflection onto the ground in the shape of an ``X''-pattern. Despite deriving from a millennia old tradition, magic mirrors inspired a measurement technique (called Makyoh topography after the Japanese word for ``wonder mirror") for detecting surface deformities in silicon wafers in the late 20th century~\cite{kugimiya1990characterization,laczik2000quantitative}. This approach has the advantage of being very simple and practical for industry based applications, in comparison with other measurement techniques such as interferometry or atomic force microscopy~\cite{kugimiya1988characterization}. 

The mathematical description and final understanding of how the magic mirrors worked was derived in 2005 by Sir Michael Berry~\cite{berry2005oriental}. Here, it was shown that the intensity of the image is given to the first order approximation by the Laplacian of the height of surface reliefs on the mirror. The principle of the magic mirror can be applied to devices working in transmission, the so called ``magic windows", which can produce a similar effect forming the Laplacian image through very slight thickness deformations~\cite{berry2017laplacian}. Specifically, the surface should be ``smooth" enough, with gentle variations, such that caustics are not formed before the image appears. It is shown that the intensity of the Laplacian image is given in terms of the height of the surface relief, $h$, by $I_\text{Laplacian~Mirror}(\textbf{r},Z) \simeq 1 + Z\,\nabla^2 h(\textbf{r})$~\cite{berry2005oriental}. Here, $Z=2D/M$ and $\textbf{r}=R/M$ are the distance along the propagation direction and transverse position from the centre of the mirror respectively, normalized to the magnification of the convex mirror $M$. $D$ and $R$ are the distance from the mirror and transverse position of the image, respectively. The Laplacian image produced from a magic window, however, depends on the relative refractive index of the window, $n$, in addition to the height of the surface relief, $h$, which is given by~\cite{berry2017laplacian},
\begin{equation}
I_\text{Laplacian~Window}(\textbf{R},z) \simeq 1 -z\,(n-1)\,\nabla^2 h(\textbf{R}).
\label{eq:1}
\end{equation}
Here $z$ is the distance of the image plane from that of the window, and $R$ is the transverse distance from the center of the window. Given any image, we can thus find the necessary surface of the magic window or mirror by solving the Poisson equation in the transverse plane.

There has been recent interest in the problem of shaping light intensity often referred to as freeform optics~\cite{brand2019freeform}. In this work we show how magic mirrors/windows can be implemented with flat optical devices. We use liquid crystal (LC) based devices, a reflective Spatial Light Modulator (SLM), and a Pancharatnam-Berry Optical Phase Element (PBOE), to construct our magic mirror and magic window~\cite{larocque2016}. The latter, through an effect known as light's spin-to-orbital angular momentum coupling~\cite{cohen2019geometric}, allows us to implement a polarization dependent phase distribution. Thus, one can observe the image resulting from the phase or its negative by switching the input polarization from left to right circular. In addition, the introduced topic of spin-to-orbit coupling allows one to explore the complex patterns of polarization singularities when the LC magic plate is illuminated with a linear superposition of left and right-handed circular polarizations. There have been many investigations into singular optics, arising from such polarization singularities, introduced in polarization system such as Stokes singularities, polarization knots and links\cite{flossmann2008polarization,larocque2018reconstructing,dennis2002polarization,cardano2013generation}. It is thus interesting to reconstruct the polarization topology of the images formed by a LC magic plate. In particular, we track the trajectory in the three-dimensional space of C-point singularities, i.e., loci of circular polarization. We show that C-points accumulate in points of the transverse plane where the image is forming.

\begin{figure}[!htb]
	\begin{center}
		\includegraphics[width=1.0\columnwidth]{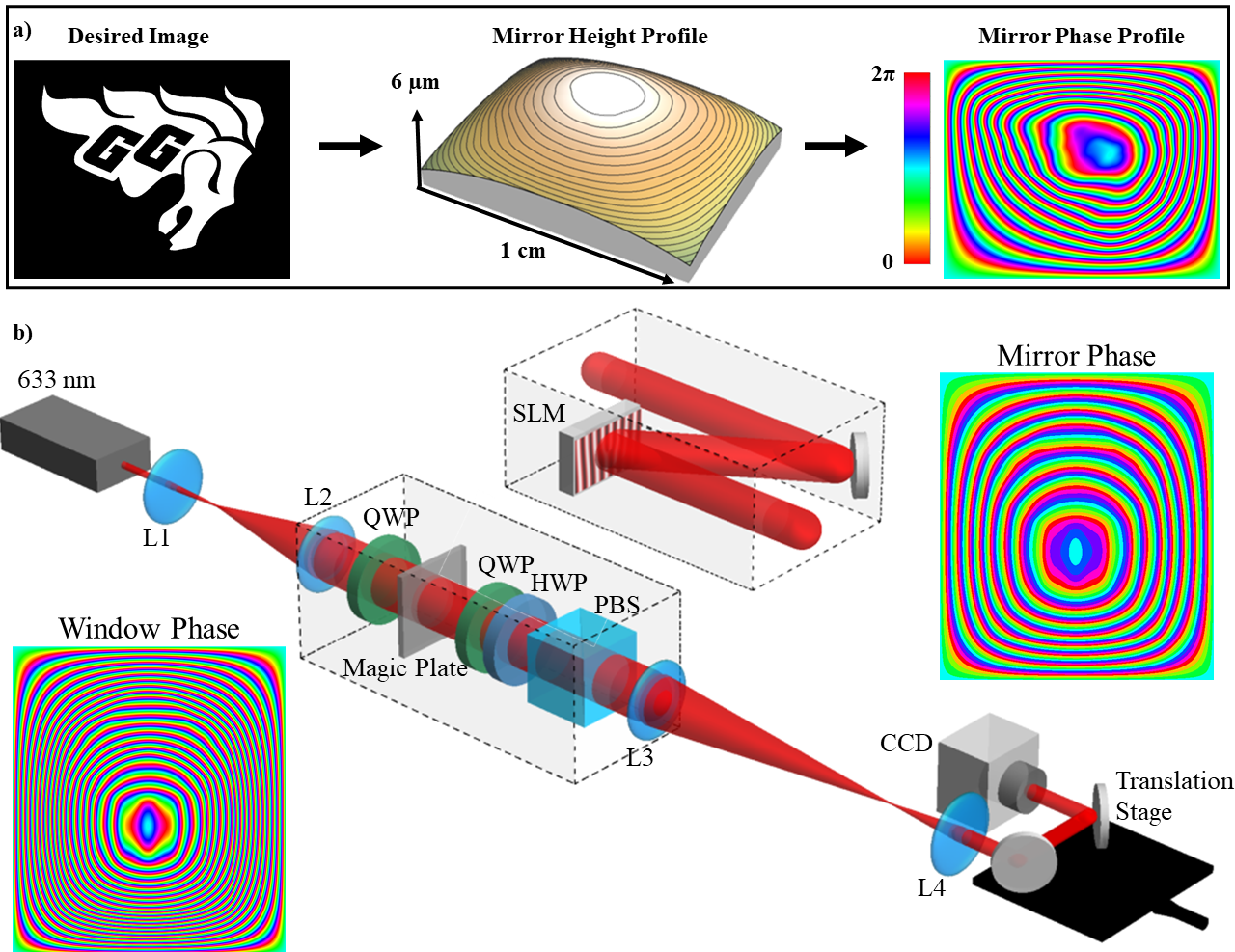}
		\caption[]{Calculation principle of phase patterns and experimental setup.
		In \textbf{a} we illustrate the steps used to calculate the phase pattern needed for generating a desired image. The BMP image of the desired intensity is used to calculate the mirror height using the discrete sine transform to solve equation \eqref{eq:1}. The maximum height for the window corresponding to the Gee-Gees logo is $6~\mu \text{m}$ as shown. The mirror height is then used to calculate the mirror phase by taking the modulus  for the specific wavelength, i.e., $\text{Phase} =\text{Mod}( \text{Height}/(2\pi\lambda),  2\pi) $. In \textbf{b} we show the experimental setup. A 633~nm He-Ne laser was used for characterizing the magic mirror and magic plate. The setup for the magic mirror \textbf{b} consists of a spatial light modulator (SLM) with a resolution of 600 by 800 pixels. After the SLM, a 4-f system is used to image the pattern displayed on the SLM, allowing for us to both make measurements starting precisely from the SLM plane and also filter out the first diffraction order using a pinhole placed at the focus of the 4-f system. Following the 4-f lens system, 2 mirrors are placed on a translation stage to construct a trombone which is followed by a CMOS camera. In the magic window setup the PBOE is placed in the same plane as the SLM. The addition of a QWP before, and a QWP, HWP, and PBS after are required to perform polarization tomography on the output beam of the magic window.}
		\label{fig:setup}
	\end{center}
\end{figure}

The goal of the experiment is to show a magic mirror and magic plate using liquid crystal technology in an SLM and PBOE, respectively. To achieve this, we must generate the phase pattern corresponding to the chosen intensity image. The image patterns that we use are converted to a bitmap form such that the pixels contain only a 1 or 0 for the intensity, see Fig.~\ref{fig:setup}-{\bf a}. Based on Eq.~\eqref{eq:1} the discretized intensity function $I(\textbf{R},z)$ is then used to find the height of the surface relief $h(\textbf{r})$, where $\mathbf{r}=\mathbf{r}(\mathbf{R})$ at the image plane. The inverse of the Laplacian is solved numerically with Dirichlet boundary conditions using the 2-dimensional discrete sine transform. In the case of flat optics, we are not varying the height of the window, but rather the index of refraction $n(\textbf{r})$. Thus, Eq.~\eqref{eq:1} becomes $I_\text{Magic Plate}(\textbf{R},z) = 1 + h\,z\, \nabla^2 \, n(\textbf{R})$, where $n(\textbf{r})$ is the transverse spatially dependent index of refraction of the plate. The resulting window phase pattern, as shown in Fig.~\ref{fig:setup}-{\bf a}, is plotted in radians.

\begin{figure}[t]
	\begin{center}
		\includegraphics[width=1.\columnwidth]{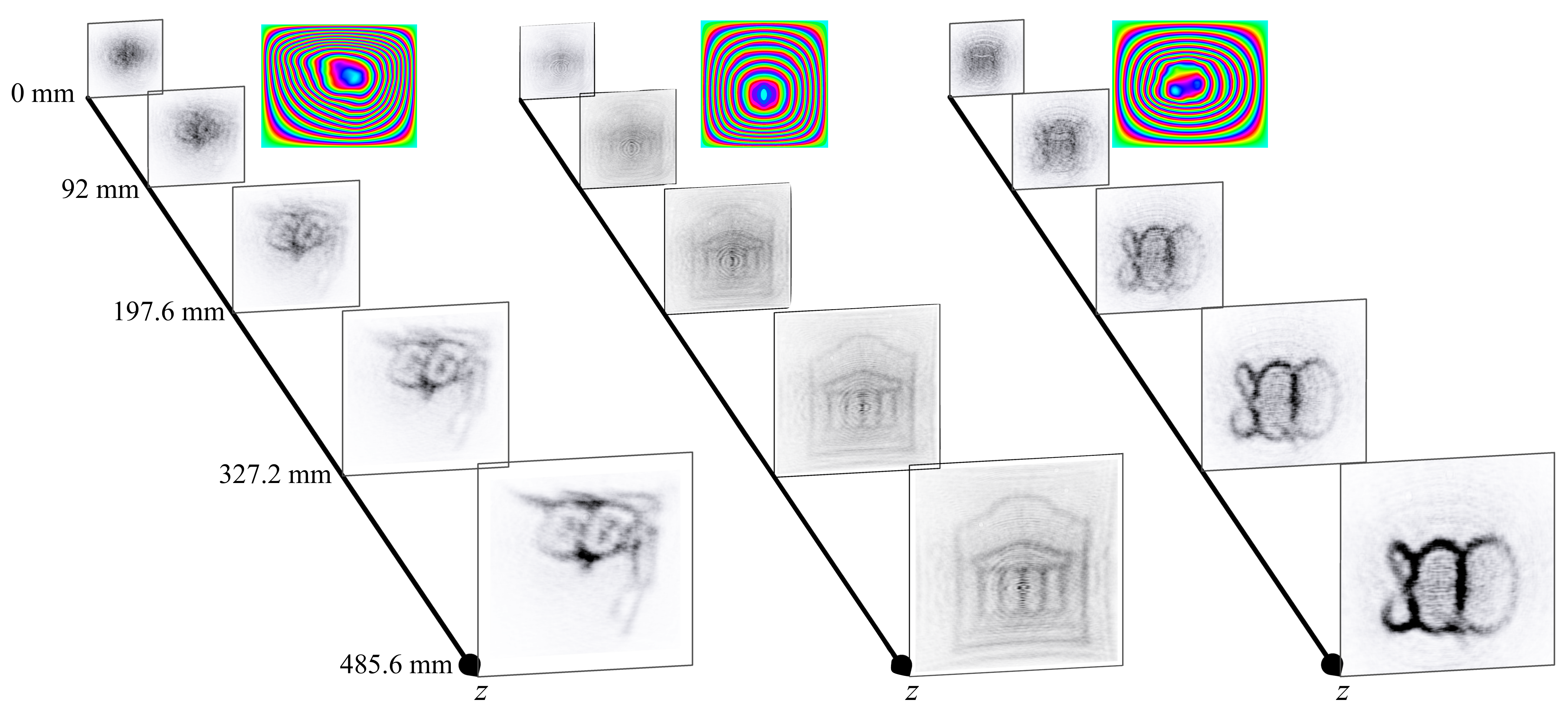}
		\caption[]{Intensity images recorded at different propagation distances after reflection from an SLM-based magic mirror. The first image at the back shows the SLM plane, where there is a uniform intensity pattern. The propagation shows the contrast of the image improving upon propagation away from the SLM. The insets show the phase distributions, for the 3 different examples, encoded on the SLM in hue colors, which encode a phase range from 0 to 2$\pi$.}
		\label{fig:SLM}
	\end{center}
\end{figure}
The magic mirror is realized using a Hamamatsu SLM with a screen resolution of 800 by 600 pixels. Figure~\ref{fig:setup}-{\bf b} shows the detail of the experimental setup for both magic mirror and magic plate. Given the desired image, the required phase pattern for the mirror is computed. There is a freedom to increase the steepness of the pattern, i.e., increasing the number of times the phase pattern goes from 0 to $2\pi$. This can be seen as altering the concavity of the mirror which results in changing how quickly the image is formed. In practical settings, care must be taken in choosing the window size and phase steepness. Due to beam divergence, an image which forms too slowly will lose its sharpness. At the same time, the pattern must be smooth enough such that caustics are not formed before the image plane~\cite{berry2017laplacian}. The phase pattern was uploaded to the SLM with the addition of a vertical grating. The first order of diffraction is selected, filtering out the rest using an iris in the center of a 4-f lens system, to remove unconverted light resulting from inefficiencies in the SLM. In addition to selecting the first order of diffraction, the 4-f lens system is also used to image the SLM plane and probe the intensity at different propagation distances. The evolution of the intensity distribution is recorded on a CMOS camera from the plane of the SLM down to the image formation planes. Additional planes were also imaged beyond the latter to capture the formation of caustics. The implementation of three SLM magic mirrors is shown in Fig.~\ref{fig:SLM}, where the intensity distribution smoothly evolves from the input beam profile to the desired image pattern.

\begin{figure*}[t]
	\begin{center}
		\includegraphics[width=2.0\columnwidth]{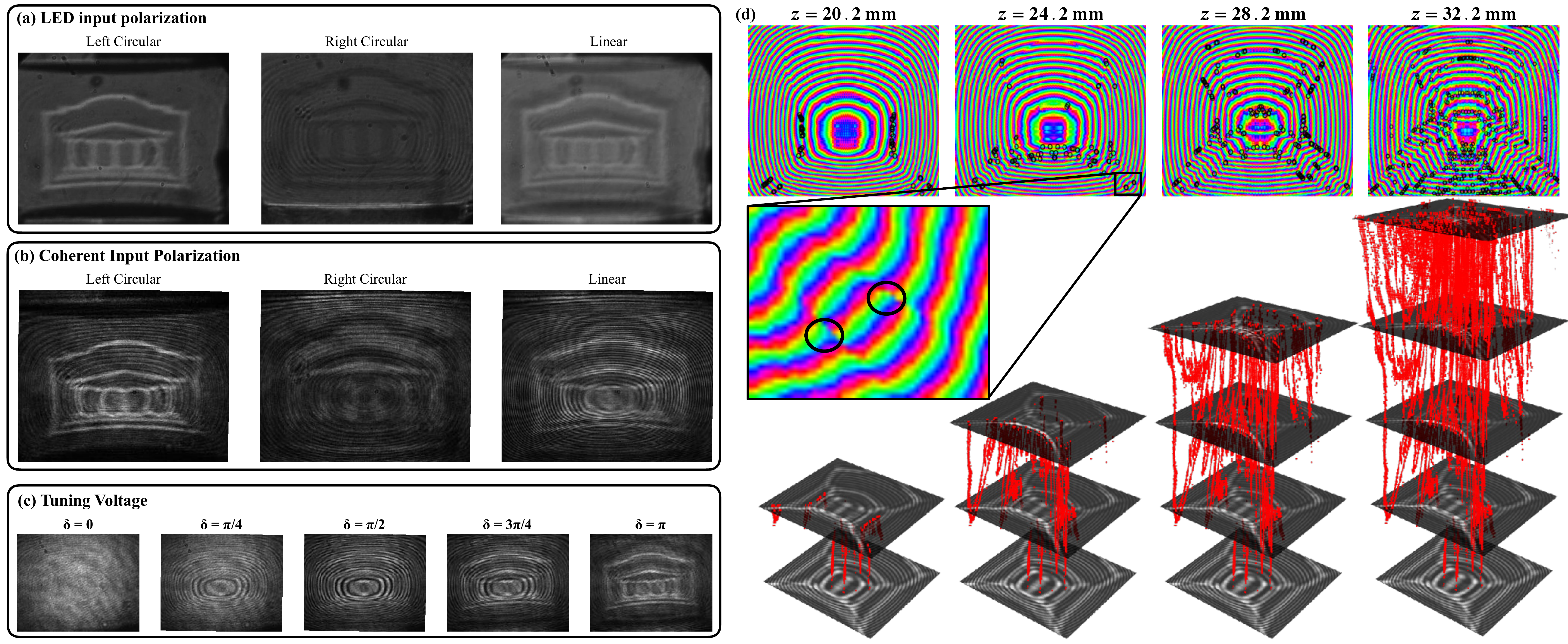}
		\caption[]{Working principle of a flat spin-orbit magic plate. In {\bf a} and {\bf b} we show the intensity distributions of light transmitted by the plate with different polarization inputs: left circular, right circular, and linear polarization from an incoherent LED source, \textbf{a}, and a 633~nm He-Ne laser, \textbf{b}. The resulting image from the linear input is the sum of the left and right circular contributions. The image resulting from the right circular input is not an exact negative of the left circular because it experiences a slight de-focusing from the window while the left component is slightly focused. Panel \textbf{c} shows the images resulting from different $\delta$ across the plate as given by Eq.~\eqref{eq:2}, with the left circular polarization input. When $\delta = 0$ there is no conversion from left to right handed polarization, and thus the beam does not acquire the phase of the magic window. When $\delta = \pi$, the input polarization is fully converted and thus we see the desired image. Tuning parameters between $0$ and $\pi$ result in partial conversion and we see interference between the intensity profile of the converted beam and the input beam. In \textbf{d}, we show the propagation of C-points from the spin-orbit magic plate. The red points in the 3-dimensional graphics show the C-points trajectory along the propagation of the beam. The intensity profiles are shown at different propagation planes. The 2-dimensional plots above represent the polarization azimuth, where C-points are labeled with black circles. The inset shows a magnified image of the polarization azimuth where the singularities are clearly visible at the location where the polarization azimuth is undefined. Upon close observation, we can see that these singularities have opposite charge. The hue color coding corresponds to polarization azimuth values ranging from 0 to $\pi$.}
		\label{fig:qplate}
	\end{center}
\end{figure*}
As the next step, we bring together the concepts of magic window imaging and photonic spin-orbit coupling. Such a device, which we call the spin-orbit magic plate, is based on PBOEs, i.e., slabs of uniaxial anisotropic materials (liquid crystals, in our case) with an extraordinary axis orientation that is spatially varying in the plate's plane. The action of a PBOE element with extraordinary axis orientation $\chi(\mathbf{r})/2$ and (spatially uniform) retardation $\delta$ is given by
\begin{equation}\label{eq:2}
\mathbf{e}_{\pm} 
\;\;
\underrightarrow{\text{MP}}\;\;
\text{cos}\left(\frac{\delta}{2}\right)
\mathbf{e}_{\pm}\, 
+\text{i}\;\text{sin}\left(\frac{\delta}{2}\right)\,
\mathbf{e}_{\mp}\, e^{\pm\, i\chi(\mathbf{r})}\,,
\end{equation}
where $\mathbf{e}_+$ and $\mathbf{e}_-$ stand for the left and right circular polarization unit vectors, respectively. The sample optical retardation, $\delta$, can be tuned by applying an AC voltage to the plate. Here, we can see that a perfectly tuned PBOE with $\delta = \pi$ results into the complete conversion of the input circular polarization to the opposite handedness with the addition of the desired phase $\pm\chi(\mathbf{r})$, where $\chi(\mathbf{r})$ is the inverse Laplacian of the image. Therefore, by flipping the incident polarization state from left to right handed, one can gain a $+\chi(\mathbf{r})$ and $-\chi(\mathbf{r})$ phase at the output, respectively.

The University of Ottawa logo was chosen to be used for the spin-orbit magic plate. The plate is fabricated in our own liquid crystal facility~\cite{larocque2016}. A pair of ITO glass plates are spin-coated with a polyamide. The ITO plates then are kept at 4~$\mu$m-distance using spacers, and glued to each other with epoxy glue. The chosen polyamide can be photoaligned through illumination from linearly polarized UV light. We are able to control the orientation of the polyamide by changing the polarization of an incident UV-beam on a pixel by pixel basis by using a digital micromirror device (DMD). The pattern written on the polyamide dictates the orientation of the liquid crystal molecules, which are added between the plates in the successive stage. The pattern was written with 32 phase steps, thus 32 polarization settings illuminating different parts of the plate. 
We characterized the action of the fabricated plate illuminating it with both incoherent and coherent light. The magic plate optical retardation was set to $\delta=\pi$ by applying an AC voltage. An incoherent source, emitted light from an LED followed by a polarizing beam splitter, was used to illuminate the magic plate. The light was filtered with a bandpass filter centered at 633~nm. We choose the input polarization to be either linear or right/left circular by using a quarter-wave plate before illuminating the magic plate. The transmitted intensity was recorded on the CMOS camera. With the linear polarized input, we observe the simultaneous formation of the University of Ottawa logo and its negative image (with an imperfect overlap due to polarization dependent lensing), as shown in Fig.~\ref{fig:qplate}-{\bf a}. The appearance of the negative image is due to the input right circular polarization component, which gains the phase $-\chi(\mathbf{r})$ (this has the effect of flipping the relative sign in Eq.~\eqref{eq:1}). It is possible to isolate the image or its negative by choosing input left or right circular polarization, respectively (Fig.~\ref{fig:qplate}-{\bf a}). 
Similar effects are observed in the case of illumination with a coherent laser beam. We used a He-Ne laser ($\lambda= 633$ nm) prepared with left/right circular or linear polarization. The resulting intensity in the case of input left circular polarization corresponds to the desired pattern (Fig.~\ref{fig:qplate}-{\bf b}). We also observe fringes due to the transverse coherence of the source. As in the incoherent illumination case, an input right circular polarization gives rise, within the magic window theory approximations, to the negative of the desired image. When a beam with linear polarization is sent onto the magic plate, as opposed to one of the circular polarizations, the resultant image is a coherent linear combination of the images one would achieve from a left and right circular input. Moreover, the magic plate optical retardation $\delta$ can be altered to not be $\pi$, but any other values. In Fig.~ \ref{fig:qplate}-{\bf c}, we show how, by tuning the optical retardation $\delta$, we can switch, at a given plane, between the input beam intensity distribution and the image encoded in the plate. 
The interplay between source coherence and polarization conditioned action of the device leads to the formation of polarization singularities during the beam propagation. The polarization of an optical field can generally be described by the ellipse traced out by the electric field vector upon propagation. The polarization ellipse can then be defined by how close the ellipse is to a circle, its ellipticity, and the orientation of the ellipses major axis, called the azimuth. The polarization azimuth is defined as the phase of $\sigma = \abs{\sigma}e^{\text{i}\theta/2} \equiv S_1 + \text{i}\,S_2 $, where $\theta$ is the angle of the polarization ellipse, and $S_1 = I_H - I_V$ and $S_2 = I_D - I_A$ are the Stokes parameters defined in terms of the measured intensity patterns of the horizontal ($I_H$), vertical ($I_V$), diagonal ($I_D$), and antidiagonal ($I_A$) polarizations, respectively~\cite{dennis2009singular}. When an electric field has a non-uniform polarization pattern, an interesting phenomenon can arise whereby the polarization azimuth is undefined. These singularities of the complex scalar field are called C-points. C-points are loci of exactly circular polarization, thus the orientation of the major axis of the polarization ellipse cannot be defined. When the magic plate is illuminated by linearly polarized light, the outgoing beam has both the right and left circular component, whose propagation is dictated by, respectively, the plate phase and the negative of the plate phase. The interaction of these two co-propagating beams gives rise to C-points with different topological charges. In the plane of the liquid crystal magic plate, the polarization remains uniformly linear, since we still have a uniform intensity distribution, albeit with a different phase for the left and right components, i.e., $(e^{+\text{i}\chi(\mathbf{r})}\,\mathbf{e}_{-}+e^{-\text{i}\chi(\mathbf{r})}\,\mathbf{e}_{+})/\sqrt{2}$. It is not until propagation to the image plane where differences in the intensity pattern of the right and left components result in the appearance of the polarization C-points. Due to conservation of total topological charge, C-points appear at given planes in pairs with opposite topological charges. The free-space dynamics of the C-points generated here is rich and requires to be investigated individually.

In summary, we have realized a liquid crystal-based magic plate exploiting the principle of light manipulation with flat optics, where the impinging light wavefront is modulated by an inhomogeneous refractive index distribution. By exploiting the physics of patterned anisotropic media, we fabricated a flat spin-orbit magic plate. This device, depending on whether the input polarization is circularly left- or right-handed, creates a desired pattern or its negative, respectively. The flat magic plate can be tuned for operation at different wavelengths since its optical retardation can be adjusted by applying an external electric field to the plate. The working principle was demonstrated for both incoherent and coherent sources. In the latter case, interference effects lead to the formation of polarization singularities (C-points) whose trajectories were experimentally reconstructed. 

\noindent E.K. acknowledges the fruitful conversation with Sir Michael Berry. This work was supported by Canada Research Chairs; Ontario Early Research Award (ERA); Canada First Research Excellence Fund (CFREF) and Natural Sciences and Engineering Research Council of Canada (NSERC).

\bibliographystyle{naturemag}
\bibliography{magicWindow}

\end{document}